# DNA to DNA Transcription Might Exist in Eukaryotic Cells

**Gao-De Li**

Chinese Acupuncture Clinic, Liverpool, UK
Email: gaode_li@yahoo.co.uk

## Abstract

**Till now, in biological sciences, the term, transcription, mainly refers to DNA to RNA transcription. But our recently published experimental findings obtained from *Plasmodium falciparum* strongly suggest the existence of DNA to DNA transcription in the genome of eukaryotic cells, which could shed some light on the functions of certain noncoding DNA in the human and other eukaryotic genomes.**

## Keywords

*Plasmodium falciparum*, Chloroquine, Genome, CAGFs, Noncoding DNA, Junk DNA, DNA to DNA Transcription, Single-Stranded DNA (ssDNA), ssDNA Transcript

**Subject Areas:** Cell Biology, Genetics, Molecular Biology

## 1. Introduction

Till now, in biological sciences, the term, transcription, mainly refers to DNA to RNA transcription, in which certain segment of DNA sequence in the genome is copied into either protein-coding RNA (mRNA), or functional non-protein-coding RNA (tRNA, rRNA, microRNA, and catalytic RNA). The DNA to RNA transcription mainly solves the problem of how a small fraction of protein-coding DNA in the genome works, but cannot solve the problem of why the large amounts of non-protein-coding DNA in the genome are non-functional.

For many decades, it has been believed that 98% of the human genome is composed of noncoding DNA which is not functional and thus named as "junk DNA" [1]. In 2012, a publication from the Encyclopedia of DNA Elements (ENCODE) project claimed that 80% of the human genome was biochemically functional [2], which was immediately followed by criticisms because the criteria used by the claim were extremely loose [3]. Recently published research results suggest that 8% - 10% of the human genome is likely functional [3] [4], which means that the large fraction of the human genome is still thought to be useless. The divergence of opinions about the functions of noncoding DNA in the human genome arises probably due to the fact that the main mechanism by which non-coding DNA exerts its functions has not yet been discovered.

Recently published our new research findings showed that *Plasmodium falciparum* could produce cell-cycle-associated amplified genomic-DNA fragments (CAGFs) during its intraerythrocytic cycle. The CAGFs are thought to be single-stranded DNA molecules which might be synthesized through the process of DNA to DNA transcription [5]. If the mechanism of DNA to DNA transcription exists in the eukaryotic cells, the problem related to the functions of noncoding DNA in the human and other eukaryotic genomes could be at least partially resolved.

Since only few words about DNA to DNA transcription were mentioned in our previous publication [5], which were not enough to address this important issue, therefore, more thoughts about DNA to DNA transcription are given in this paper.

## 2. DNA to DNA Transcription Is a Hypothesis

Based on our published experimental findings we hypothesize that in addition to DNA to RNA transcription, there is a mechanism of DNA to DNA transcription in the genome of eukaryotic cells, which involves using DNA polymerase to copy particular segment of noncoding DNA sequence in the genome into a single-stranded DNA (ssDNA) molecule that can be named as ssDNA transcript. The ssDNA transcripts are the same molecules as CAGFs reported in our precious publication, and thus have the same functions as CAGFs, *i.e.*, being involved in the regulation of gene expression during cell cycle progression, and generating genetic alterations in the eukaryotic cells that are exposed to prolonged environmental stress [5] [6].

Since the DNA sequence by which a protein-coding RNA (mRNA) is transcribed is called a protein encoding gene, and the DNA sequence by which a functional non-protein-coding RNA (tRNA, rRNA, etc.) is transcribed is called an RNA gene, the DNA sequence by which a functional non-protein-coding ssDNA molecule is transcribed therefore should be called a DNA gene. From both protein encoding gene's and RNA gene's point of view, the large amounts of noncoding DNA in the human and other eukaryotic genomes are useless, but from the DNA gene's point of view, these large amounts of noncoding DNA could be very useful as they might harbour many unknown DNA genes which might be required for cell cycle regulation and cellular stress responses [6] [7].

## 3. Discovery of CAGFs Is a Strong Support for the DNA to DNA Transcription Hypothesis

Our recently published experimental findings showed that *P. falciparum* could produce CAGFs at different points of its intraerythrocytic cycle, and that chloroquine (CQ) was found to induce the production of CAGFs in the malaria parasite [5]. As for the nature of CAGFs, we speculate that CAGFs are ssDNA molecules, which is based on the following reasons.

Since CAGFs are amplified (multiple copies are produced) and degraded during cell cycle progression, they are certainly not incorporated into the genome, but released from the genome into the nucleoplasm. Furthermore, CAGFs are amplified before cell cycle progression into DNA synthesis phase (S phase) of *P. falciparum*, which indicates that an unknown DNA synthesis mechanism might exist outside S phase of the cell cycle. What is this unknown DNA synthesis mechanism? Is it the same as the semiconservative mode of DNA replication in S phase? The answer is certainly no because DNA replication in S phase refers to the whole genome duplication, even if the DNA replication mechanism also works outside of S phase to duplicate DNA at certain regions of the genome, it is impossible to release these duplicated DNA regions as double-stranded DNA fragments from the genome without affecting the genome integrity.

Therefore, it is reasonable to postulate that CAGFs are ssDNA molecules which might be synthesized in the way similar to RNA, but using DNA polymerase instead of RNA polymerase during the process of transcription. The process could be named as DNA to DNA transcription, and CAGFs could be thought as ssDNA transcripts. If CAGFs are ssDNA molecules transcribed from some noncoding DNA in the genome, nearly all the features of CAGFs, such as genome-released, multicopy-produced, drug-inducible and easily degraded during the cell cycle progression could be reasonably explained. Besides, if CAGFs are ssDNA molecules transcribed from noncoding DNA in the genome, their role in the regulation of gene expression during cell cycle progression, and generating genetic alterations in the eukaryotic cells will be easily understandable as they can bind to complementary ssDNA regions at transcription sites. In a word, discovery of CAGFs is a solid evidence to support the DNA to DNA transcription hypothesis.

## 4. Ideas for Further Experimental Test of the Hypothesis

Theoretically, if DNA to DNA transcription exists in eukaryotic cells, it is possible to isolate the ssDNA transcripts from the cells. But since we don't know the specific sequence feature related to the ssDNA transcripts, it is therefore impossible to develop a protocol to specifically isolate them from the cells. Besides, other ssDNA in nuclear DNA samples might interfere with isolation of the ssDNA transcripts. It was reported that ssDNA accounts for about 1% - 2% of total nuclear DNA isolated from various eukaryotic cells [8]-[10], which was hypothesized to be generated by selective endogenous nuclease attacks at an early stage of the DNA purification procedure[10].

At present, a simple and practical approach to proving existence of ssDNA transcripts during the eukaryotic cell cycle is to repeat our arbitrarily-primed PCR results with the PCR using the two specific primers (SP1: 5'-AATAAGATGTGCTGTTTAACT-3' and SP2: 5'-GTTGTGCTCACACTTATCT-3'), the origin of which was reported in our previous publication [5]. The PCR band produced by these two primers was called UB1-re- lated band in our previous publication [5], representing the DNA fragment of one of CAGFs. The process of this PCR-based approach is outlined below.

To repeat our previous experiments, only 25-nM dose of chloroquine (CQ) is needed to treat *P. falciparum* K1 and HB3 isolates as the dose is proved by our previous study to be the best one for inducing CAGFs. The genomic DNA from both K1 and HB3 isolates is isolated at same time points (2 h, 6 h, 12 h and 24 h) following CQ treatment, and then subject to the PCR using SP1 and SP2. The expected results will be consistent with those of UB1 reported in our previous publication [5], *i.e.*, only one clear PCR band (UB1-related band) will be found in all samples, but a very bigger PCR band which indicates amplified templates for UB1 will be found in the 12 h-isolated sample from the control groups of both K1 and HB3 isolates; in CQ-treated groups, the very bigger PCR band will be found in 2 h, 6 h and 24 h isolated samples of K1 isolates and in 2 h, 6 h 12 h and 24 h isolated samples of HB3 isolates. For relative quantification of the PCR bands, measurement of the band density can be carried out with a densitometer.

In order to check if UB1- related one of CAGFs is a ssDNA molecule, S1 nuclease can be used to treat genomic DNA samples before performing above PCR experiments. If the very bigger PCR band as described above becomes smaller in the S1 nuclease-treated groups and looks nearly the same as that in the control groups, it means that the amplified ssDNA templates for UB1 have been degraded by S1 nuclease, and thus the single-strandedness of UB1-related one of CAGFs will be confirmed. Taken together, if UB1-related one of CAGFs is proven to be single-stranded, extragenomic, CQ-inducible, and "amplified" (*i.e.*, many copies are produced) at different points of the cell cycle, it is definitely the ssDNA molecule produced by DNA to DNA transcription, and therefore the hypothesis is experimentally validated.

In this study, other techniques could also be used. Northern blotting analysis could be applied using PCR product generated by SP1 and SP2 as template for preparing radioactive probe [11] [12]. This method is very useful as it could not only detect if UB1-related one of CAGFs is amplified or not, but could also show the size of UB1-related one of CAGFs. For subcellular localization of UB1-related one of CAGFs, electron microscopy in situ hybridization could be used [13].

Since the *P. falciparum* Genome Sequencing Project has been completed [14], it is possible to obtain the whole sequence of UB1-related ssDNA transcript by checking related DNA sequences within the database. If special sequence features in the whole ssDNA transcript could be found through computer analysis, more candidate ssDNA transcripts could be picked up from the genome database of *P. falciparum*. Experimental test of these candidate ssDNA transcripts can be performed by PCR using primers that are designed according to the DNA sequences of candidate ssDNA transcripts picked up from the database.

## 5. Implications of the Hypothesis

The importance of the hypothesis is that DNA to DNA transcription is a fundamental mechanism of eukaryotic cells, which is involved in a wide-range of cellular functions. To date, the functions of noncoding DNA in the human and other eukaryotic genomes remain largely unknown, which has hampered progress in genome research. If the hypothesis is proven to be true, it could at least partially resolve this most perplexing problem, and will hugely facilitate the development of biological and life sciences.

## 6. Conclusion

DNA to DNA transcription might exist in eukaryotic cells, which is a promising hypothesis proposed based on the discovery of cell-cycle-associated amplified genomic-DNA fragments (CAGFs) in *P. falciparum*. There are many reasons to think that CAGFs are ssDNA molecules which might be produced exclusively through DNA to DNA transcription. Once the existence of DNA to DNA transcription in eukaryotic cells is confirmed, the functions of large amounts of noncoding DNA in the human and other eukaryotic genomes could be at least partially resolved. Therefore, further experimental validation of this mechanism is urgently needed.

## Conflict of Interest

The author declares that there is no conflict of interest regarding the publication of this paper.

## References

[1] Ohno, S. (1972) So Much ''Junk'' DNA in Our Genome. In: Smith, H.H., Ed., *Evolution of Genetic Systems*, Gordon and Breach, New York, 366-370.

[2] ENCODE Project Consortium (2012) An Integrated Encyclopedia of DNA Elements in the Human Genome. *Nature*, **489**, 57-74. http://dx.doi.org/10.1038/nature11247

[3] Palazzo, A.F. and Gregory, T.R. (2014) The Case for Junk DNA. *PLoS Genet*, **10**, e1004351. http://dx.doi.org/10.1371/journal.pgen.1004351

[4] Rands, C.M., Meader, S., Ponting, C.P. and Lunte, G. (2014) 8.2% of the Human Genome Is Constrained: Variation in Rates of Turnover across Functional Element Classes in the Human Lineage. *PLoS Genet*, **10**, e1004525. http://dx.doi.org/10.1371/journal.pgen.1004525

[5] Li, G.D. (2016) Certain Amplified Genomic-DNA Fragments (AGFs) May Be Involved in Cell Cycle Progression and Chloroquine Is Found to Induce the Production of Cell-Cycle-Associated AGFs (CAGFs) in *Plasmodium falciparum*. *Open Access Library Journal*, **3**, e2447. http://dx.doi.org/10.4236/oalib.1102447

[6] Li, G.D. (2016) "Natural Site-Directed Mutagenesis" Might Exist in Eukaryotic Cells. *Open Access Library Journal*, **3**, e2595. http://dx.doi.org/10.4236/oalib.1102595

[7] Galhardo, R.S., Hastings, P.J. and Rosenberg, S.M. (2007) Mutation as a Stress Response and the Regulation of Evolvability. *Critical Reviews in Biochemistry and Molecular Biology*, **42**, 399-435. http://dx.doi.org/10.1080/10409230701648502

[8] Hanania, N., Schaool, D., Poncy, C., Tapiero, H. and Harel, J. (1977) Isolation of Single Stranded Transcription Sites from Human Nuclear DNA. *Cell Biology International Reports*, **1**, 309-315.

[9] Leibovitch, S.A. and Harel, J. (1978) Active DNA Transcription Sites Released from the Genome of Normal Embryonic Chicken Cells. *Nucleic Acids Research*, **5**, 777-787. http://dx.doi.org/10.1093/nar/5.3.777

[10] Hanania, N., Shaool, D. and Harel, J. (1982) Isolation of a Mouse DNA Fraction Which Encodes More Informational Than Non Informational RNA Sequences. *Molecular Biology Reports*, **8**, 91-96. http://dx.doi.org/10.1007/BF00778510

[11] Alwine, J.C., Kemp, D.J. and Stark, G.R. (1977). Method for Detection of Specific RNAs in Agarose Gels by Transfer to Diazobenzyloxymethyl-Paper and Hybridization with DNA Probes. *Proceedings of the National Academy of Sciences of the United States of America*, **74**, 5350-5354. http://dx.doi.org/10.1073/pnas.74.12.5350

[12] Kevil, C.G., Walsh, L., Laroux, F.S., Kalogeris, T., Grisham, M.B. and Alexander, J.S. (1997) An Improved, Rapid Northern Protocol. *Biochemical and Biophysical Research Communications*, **238**, 277-279.

[13] Cmarko, D. and Koberna, K. (2007) Electron Microscopy in Situ Hybridization: Tracking of DNA and RNA Sequences at High Resolution. In: *Methods in Molecular Biology™*, **369**, 213-228. http://dx.doi.org/10.1007/978-1-59745-294-6_11

[14] Gardner, M.J., Hall, N., Fung, E., White, O., Berriman, M., Hyman, R.W., *et al*. (2002) Genome Sequence of the Human Malaria Parasite *Plasmodium falciparum*. *Nature*, **419**, 498-511. http://dx.doi.org/10.1038/nature01097